\def \AAP #1 #2 {{\em Astron. Astrophys.\/} {\bf #1}, #2}
\def \AAL #1 #2 {{\em Astron. Astrophys. Lett.\/} {\bf #1}, L#2}
\def \AAR #1 #2 {{\em Astron. Astrophys. Rev.\/} {\bf #1}, #2}
\def \AAS #1 #2 {{\em Astron. Astrophys. Suppl. Ser.\/} {\bf #1}, #2}
\def \AJ #1 #2 {{\em Astron. J.\/} {\bf #1}, #2}
\def \ANNREV #1 #2 {{\em Ann. Rev. Astron. Astrophys.\/} {\bf #1}, #2}
\def \APJ #1 #2 {{\em Astrophys. J.\/} {\bf #1}, #2}
\def \APJL #1 #2 {{\em Astrophys. J. Lett.\/} {\bf #1}, L#2}
\def \APJS #1 #2 {{\em Astrophys. J. Suppl.\/} {\bf #1}, #2}
\def \APSS #1 #2 {{\em Astrophys. Space Sci.\/} {\bf #1}, #2}
\def \ASR #1 #2 {{\em Adv. Space Res.\/} {\bf #1}, #2}
\def \BAIC #1 #2 {{\em Bull. Astron. Inst. Czechosl.\/} {\bf #1}, #2}
\def \JSQRT #1 #2 {{\em J. Quant. Spectrosc. Radiat. Transfer\/} {\bf #1}, #2}
\def \MN #1 #2 {{\em Mon. Not. R. Astr. Soc.\/} {\bf #1}, #2}
\def \MEM #1 #2 {{\em Mem. R. Astr. Soc.\/} {\bf #1}, #2}
\def \PLR #1 #2 {{\em Phys. Lett. Rev.\/} {\bf #1}, #2}
\def \PASJ #1 #2 {{\em Publ. Astron. Soc. Japan\/} {\bf #1}, #2}
\def \PASP #1 #2 {{\em Publ. Astr. Soc. Pacific\/} {\bf #1}, #2}
\def \NAT #1 #2 {{\em Nature\/} {\bf #1}, #2}
\def \SAIT #1 #2 {{\em Mem.\ Soc.\ Astron.\ It.\/} {\bf #1}, #2}
\def \MESS #1 #2 {{\em The Messenger\/} {\bf #1}, #2}
\def \ASTRNACH #1 #2 {{\em Astron. Nach.\/} {\bf #1}, #2}
\def \NAR #1 #2 {{\em New Astron. Rev.\/} {\bf #1}, #2}
\def \EA #1 #2 {{\em Exp. Astron.\/} {\bf #1}, #2}

\documentstyle{BSAXPwork}
\input epsf.sty
\begin{opening}
\title{A stroboscopic method for phase resolved pulsar observations}
\author{S. Vidrih$^1$, A. \v{C}ade\v{z}$^1$, A. Carrami\~nana$^2$}
\institute{$^1$Faculty of Mathematics and Physics, University of Ljubljana,
Slovenia\\
$^2$Instituto Nacional de Astrofisica, \'Optica y Electr\'onica, M\'exico}
\date{} 
\end{opening}

\begin{document}

\oddpagefooter{}{}{} 
\evenpagefooter{}{}{} 
\medskip  

\begin{abstract}
We present a stroboscopic system developed for optical observations of pulsars
and its application in the CLYPOS survey. The stroboscopic device is connected
to a GPS clock and provides absolute timing to the stroboscopic shutter relative
to the pulsar's radio ephemerides. By changing the phase we can examine the
pulsar's light curve. The precisely timed stroboscope in front of the CCD camera
can perform highly accurate time resolved pulsar photometry and offers the
advantages of CCD cameras, which are high quantum efficiency as well as
relatively large field of view, which is important for flux calibrations.

CLYPOS (Cananea Ljubljana Young Pulsar Optical Survey) is an extensive search
for optical counterparts of about 30 northern hemisphere radio pulsars. It is a
collaboration between the INAOE, Mexico and the Faculty of Mathematics and
Physics of the University of Ljubljana. Stroboscopic observations were done
between December 1998 and November 2000 at the 2.12 m telescope of the
Observatory Guillermo Haro in Cananea, Sonora. The first results of the survey
are presented. Analyzed data indicate that there is no optical counterpart
brighter than $V\sim 22$.
\end{abstract}

\medskip

\section{Introduction}
Today more than 1000 radio pulsars are known. The interest for detection of
their optical counterparts has increased in the last decade, mainly due to
bigger and better telescopes and due to improved detectors with high temporal
resolution (Fordham et al. 2000, Dhillon et al. 2001, Moon et al. 2001,
Straubmeier et al. 2001). Despite all efforts the number
of detected optical counterparts has enlarged only to nine (Mignani et al. 2000,
Mignani et al. 2002). Almost all of them are weak optical sources,
with a $V$ magnitude of $\geq 24$. Moreover, until now there has been no
systematic search for optical counterparts of young radio pulsars. The Clypos
survey, a mexican-slovene collaboration project, was an attempt to
systematically search for optical counterparts of $\sim 30$ radio pulsars. The
goal of this survey was to detect new optical pulsars or at least to determine
magnitude limits for their detection and thus set stronger observational
constraints on theoretical high-energy emission models for pulsars.

\section{Stroboscopic observations}
Periodic optical signals from pulsars can be detected only with a device
that has sufficient temporal resolution on subsecond or even millisecond time
scale. Pulsar photometry is usually performed with high-speed photometers
(Chakrabarty et al. 1998, Golden et al. 2000). Recently the use of frame
transfer technique has also increased the temporal resolution of CCD cameras
(Kern et al. 2002, Fordham et al. 2003). We introduced the use of a precisely
timed stroboscope in front of a normal CCD camera (\v{C}ade\v{z} et al. 1996).
The stroboscopic technique is based on a shutter that opens with the prescribed
frequency and phase and thus controls when the light beam from a telescope
passes to a CCD detector. The advantages of this method are that it makes full
use of the high quantum efficiency and the detailed field of the CCD camera,
which enables good flux calibrations.

Our stroboscopic system (Kotar et al. 2002) is schematically shown in Figure 1.
The stroboscopic shutter is a rotating blade with out-cuts whose width
determines a duty cycle of the stroboscope, that is the fraction of the period
during which a CCD camera is illuminated. While observing, the rotating
frequency of the blade is synchronized with the pulsar's frequency, calculated
from published radio ephemerides (Taylor et al. 1993).
The GPS clock timing can also be used to synchronize the phase of the blade with
the absolute radio phase of the pulsar, if it is known in advance.

\begin{figure}[h]
\epsfysize=5cm 
\hspace{1.0cm} \vspace{0.0cm}
 \epsfbox{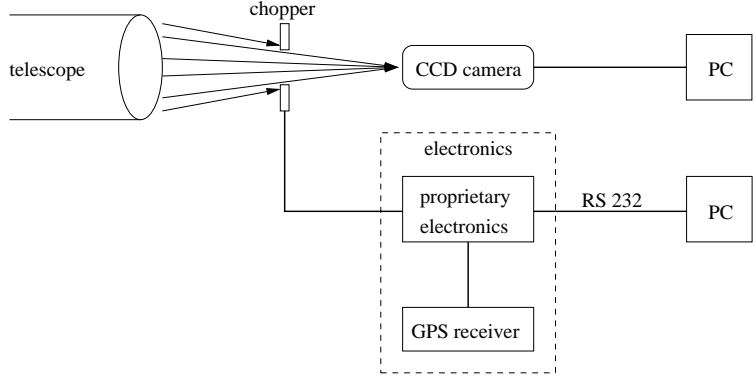}
\caption[h]{The drawing of the stroboscopic system. The CCD camera is
independently connected to another PC.}
\end{figure}

In our survey the duty cycle of the stroboscope was set to $10$\% and to $25$\%
in only a few observation runs. This decision was made under the assumption that
optical pulsar light curves are similar to the one of the Crab. The exact window
function of the stroboscope depends on the width of the blade out-cut and the
position of the blade with respect to the focal plane of the telescope. The
window function together with the Crab pulsar pulse shape obtained by Fordham et
al. (2003) is schematically shown in Figure 2a. The result of the fit of our
data to data of Fordham et al. is shown in Figure 2b.

\begin{figure}[h]
\epsfysize=4cm 
\hspace{0.0cm} \vspace{0.0cm}
 \epsfbox{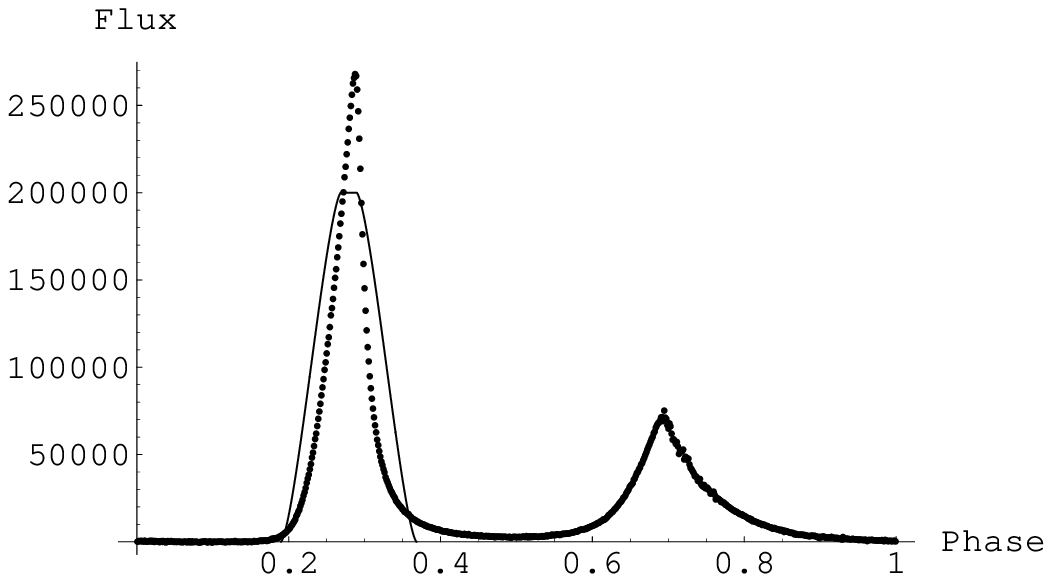}
\epsfysize=4cm 
\hspace{0.0cm} \vspace{0.0cm}
 \epsfbox{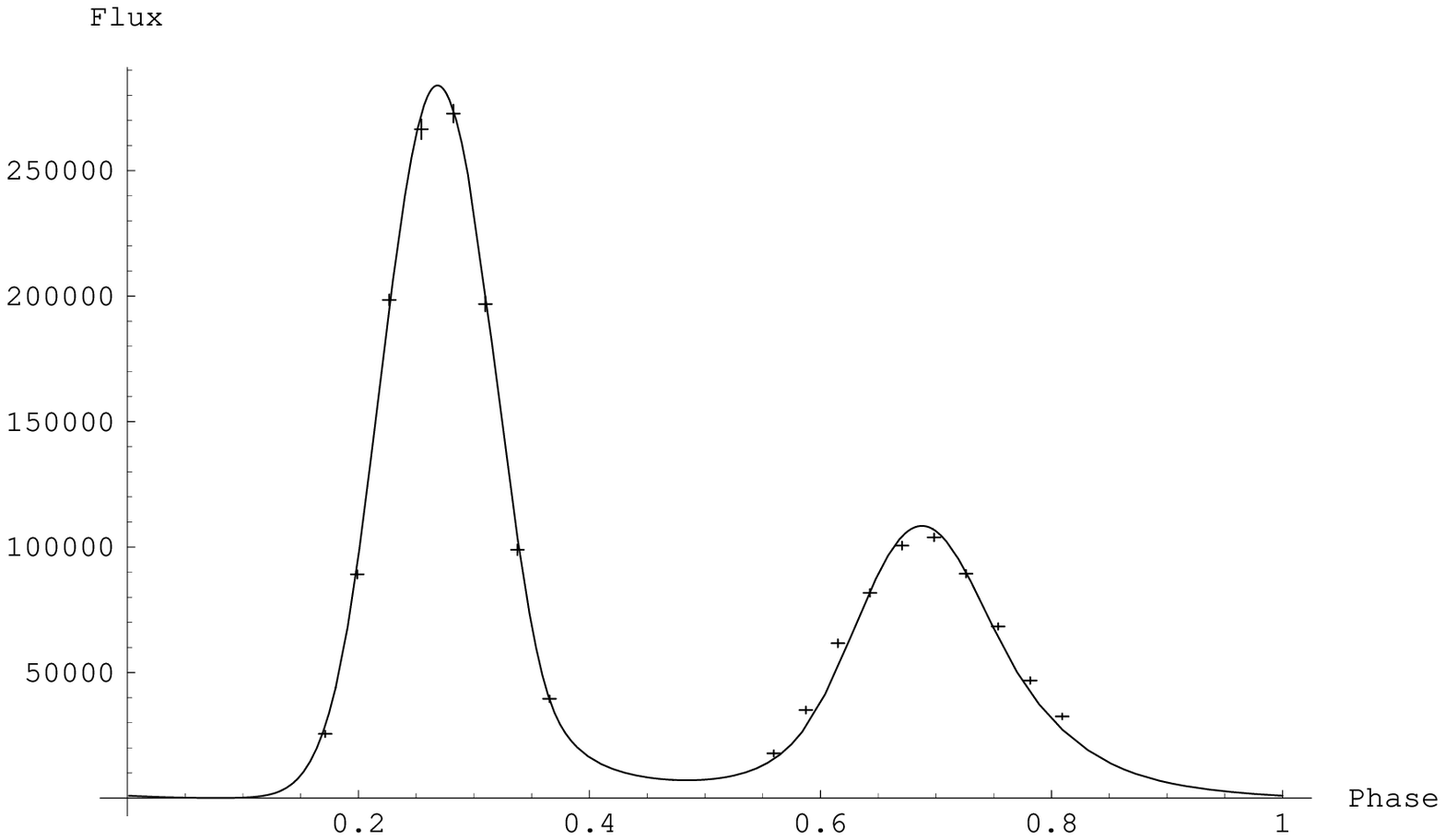}
\caption[h]{\em Left \em: The light curve of the Crab pulsar, obtained by
Fordham et al. (2003). The width of the main pulse is $\sim 10$\% of the whole
period. The calculated window function of the stroboscope is also displayed. \em
Right \em: Fit of the stroboscopic phase light curve (shown as crosses) to data
obtained by Fordham et al. and convolved with our window function (shown as a
smooth curve). In finding the fit the absolute phase and magnitude shift were
considered as free parameters.
}
\end{figure}

Knowing the stroboscopic window function we can calculate (at least for the Crab
pulsar) the magnitude of the pulsar as a function of the stroboscopic phase. The
result is shown in Figure 3, where the magnitude scale is gauged to the average
magnitude of the pulsar. The maximum magnitude enhancement of 1.8 magnitude is
reached, when the stroboscope window is completely centered on the emission
peak. One should note that broader main emission peak results in a smaller
magnitude enhancement.

\begin{figure}[ht]
\epsfysize=5cm 
\hspace{2.1cm} \vspace{0.5cm}
 \epsfbox{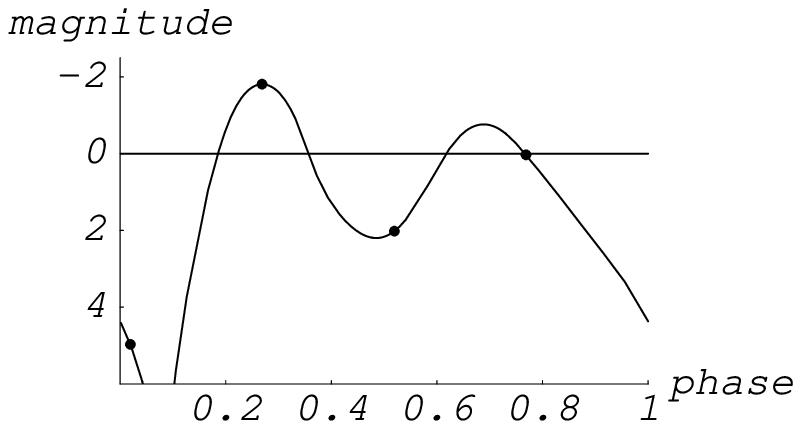}\\
\epsfysize=2.0cm 
\hspace{0.0cm} \vspace{0.0cm}
 \epsfbox{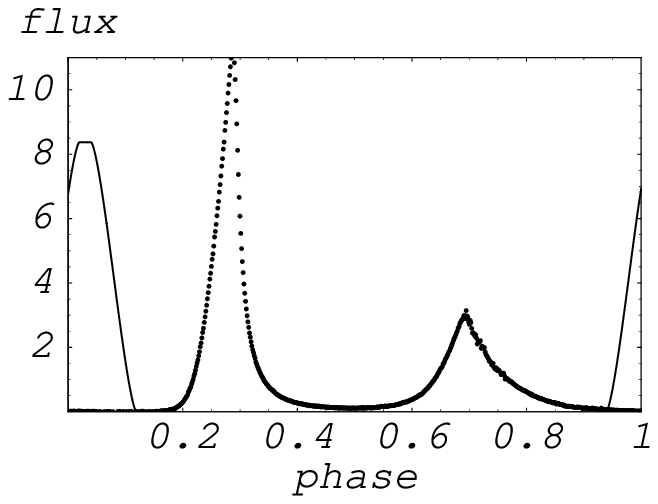}
\epsfysize=2.0cm 
\hspace{0.0cm} \vspace{0.0cm}
 \epsfbox{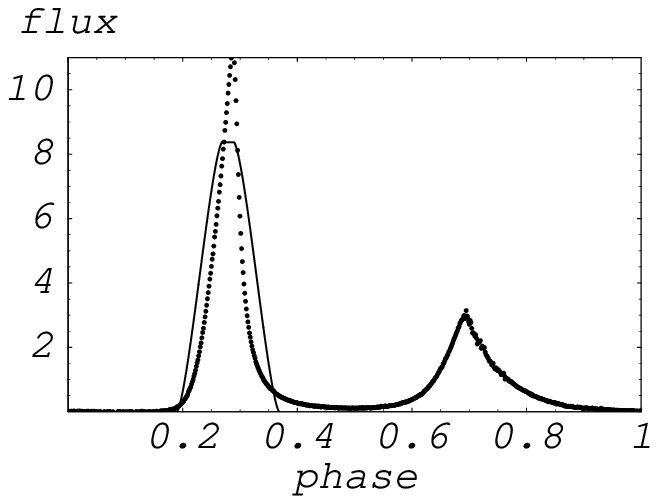}
\epsfysize=2.0cm 
\hspace{0.0cm} \vspace{0.0cm}
 \epsfbox{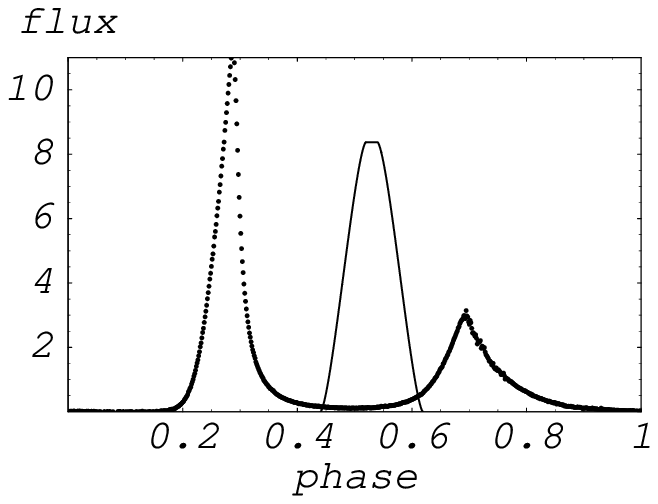}
\epsfysize=2.0cm 
\hspace{0.0cm} \vspace{0.0cm}
 \epsfbox{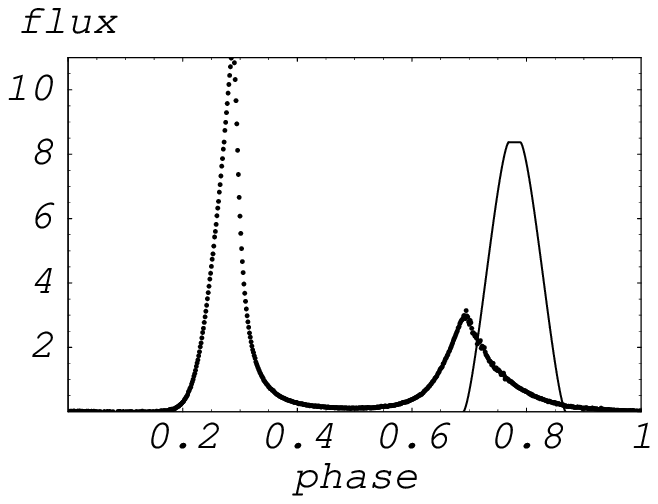}
\caption[h]{\em Up \em: Magnitude enhancement of the stroboscopic magnitude with
respect to the average magnitude of the Crab pulsar as a function of the phase
position of the stroboscope window. \em Down \em: Four positions of the
stroboscope window 1/4 of the phase apart from each other. Magnitude
enhancements/diminutions of these positions are shown as dots on the upper curve. The
maximum enhancement is 1.8 magnitude.
}
\end{figure}

\section{Clypos survey}
27 young isolated radio pulsars that were observable from Cananea (its
geographic latitude is $\sim 31^{\circ}$) were included in the Clypos survey. We
present the preliminary results for 12 of them. We were looking for the
brightest possible candidates, so our choosing criteria was the pulsar's
rotation power $\dot{E}_{rot} \propto \omega\,\dot{\omega}$. Assuming that the
optical efficiency of pulsars is the same as that of the Crab, we obtained the
expected magnitudes of optical counterparts as:
$$
V=V_C - 2.5\, \log{\left[\frac{\dot{P}/P^3}{\dot{P}_C/P_C^3}
\left(\frac{d_C}{d}\right)^2\right]}\ ,
\eqno(1)
$$
where we used the $V_C$, $P_C$, $\dot{P}_C$ and $d_C$ values for the Crab (16.5,
33.4 ms, $4.2\times 10^{-13}$, 2.2 pc). The results for the sample of 12 pulsars are shown in Table I.

\begin{table}[h]
\caption{Characteristics for 12 radio pulsars monitored in Clypos survey. The
rotation power, $\omega\,\dot{\omega}/ \omega_C\,\dot{\omega}_C$ is presented in
units of the Crab pulsar. We calculated the expected magnitude assuming that
chosen pulsars are Crab like.}
\begin{center}
\begin{tabular}{|l|r @{.} l|r @{.} l|r @{.} l||r @{.} l|}
\hline
Name & \multicolumn{2}{c|}{$\omega\,[s^{-1}]$} &
\multicolumn{2}{c|}{$\dot{\omega}\,[s^{-2}]$} & 
\multicolumn{2}{c||}{$\omega\,\dot{\omega}/ \omega_C\,\dot{\omega}_C$} &
\multicolumn{2}{c|}{V} \\\hline
\hline
PSR J 0056+4756 & 13&311 & -0&101$\times 10^{-12}$ & 3&00$\times 10^{-6}$ & 28&8
\\ \hline
PSR J 0108-1431 & 7&780 & -0&001$\times 10^{-12}$ & 0&01$\times 10^{-6}$ &
29&6\\ \hline
PSR J 0117+5914 & 61&941 & -3&570$\times 10^{-12}$ & 495&88$\times 10^{-6}$ &
\bf{24}&\bf{9} \\ \hline
PSR J 0157+6212 & 2&672 & -0&215$\times 10^{-12}$ & 1&29$\times 10^{-6}$ & 30&7
\\ \hline
PSR J 0454+5543 & 18&440 & -0&128$\times 10^{-12}$ & 5&30$\times 10^{-6}$ & 27&7
\\ \hline
PSR J 0538+2817 & 43&890 & -1&124$\times 10^{-12}$ & 110&60$\times 10^{-6}$ &
\bf{26}&\bf{1} \\ \hline
PSR J 0614+2229 & 18&760 & -3&340$\times 10^{-12}$ & 140&52$\times 10^{-6}$ &
28&0 \\ \hline
PSR J 0631+1036 & 21&836 & -7&939$\times 10^{-12}$ & 388&80$\times 10^{-6}$ &
27&6 \\ \hline
PSR J 1705-1906 & 21&015 & -0&291$\times 10^{-12}$ & 13&71$\times 10^{-6}$ &
27&5 \\ \hline
PSR J 1825-0935 & 8&171 & -0&556$\times 10^{-12}$ & 10&20$\times 10^{-6}$ & 27&5
\\ \hline
PSR J 1833-0827 & 73&675 & -7&922$\times 10^{-12}$ & 1308&91$\times 10^{-6}$ &
\bf{26}&\bf{0} \\ \hline
PSR J 1908+0734 & 29&588 & -0&115$\times 10^{-12}$ & 7&63$\times 10^{-6}$ & 30&4
\\ \hline
\end{tabular}
\end{center} 
\end{table}

Observations were carried out at the 2.12 m telescope of the Observatory
Guillermo Haro in Cananea during several observation runs between December 1998
and November 2000. The field of view of the EEV P8603 CCD detector, using the
LFOSC focal reducer ($f/2.4$), was $6 \times 10$ arcmin, with the size of one
pixel $\sim 1$ arcsec. No filters were used to maximize the light input. The
observing strategy was as follows. The exposure time for each image was 5 min,
effectively 30 sec due to the 10\% stroboscopic duty cycle. For each pulsar 24
consecutive images 1/8 in phase apart were taken. The whole phase cycle was thus
scanned 3 times. In this way we spent $\sim 2$ hours per pulsar, 
some pulsars were observed during several observation runs. A detected pulsar
would show up as a periodic signal at the radio position of the pulsar. The
brightest magnitude at the most favorable phase would be enhanced by at most
$\sim 2$ magnitudes with respect to the average magnitude. Since the absolute
radio phase of observed pulsars was not known the starting stroboscope phase was
chosen randomly.

\subsection{Limiting magnitude}
It is important to set the limiting magnitude for pulsar detection. It depends
on many factors. The average sky brightness in Cananea is 20.5 
$V/\mathrm{arcsec^2}$ and the average FWHM of the star PSF is around 3 arcsec.
The quantum efficiency of the CCD camera at the wavelength of 700 nm is 45\%.
The effective exposure time per image was 30 s and 3 images were taken with each
phase. For each pulsar field we took one V filter image in order to calibrate
the brightness of stars in the image. The magnitudes were calibrated against the
gsc2 catalog and also using photometric standards from Landolt (1992). The
limiting magnitude was determined statistically as follows. We checked how often
any star on 24 images was detected. Two examples of such detection efficiency
are shown in Figure 4. The efficiency is very close to 100\% for bright stars
and it suddenly drops for stars with magnitudes higher than a certain value
which we interpret as magnitude limit. The limiting magnitude varies for
different observation runs between $V_{min}=20$ and $V_{max}=22$, due to
changing sky conditions that mainly affect the width of the PSF.

\begin{figure}
\epsfysize=4.8cm 
\hspace{0.0cm} \vspace{0.0cm}
 \epsfbox{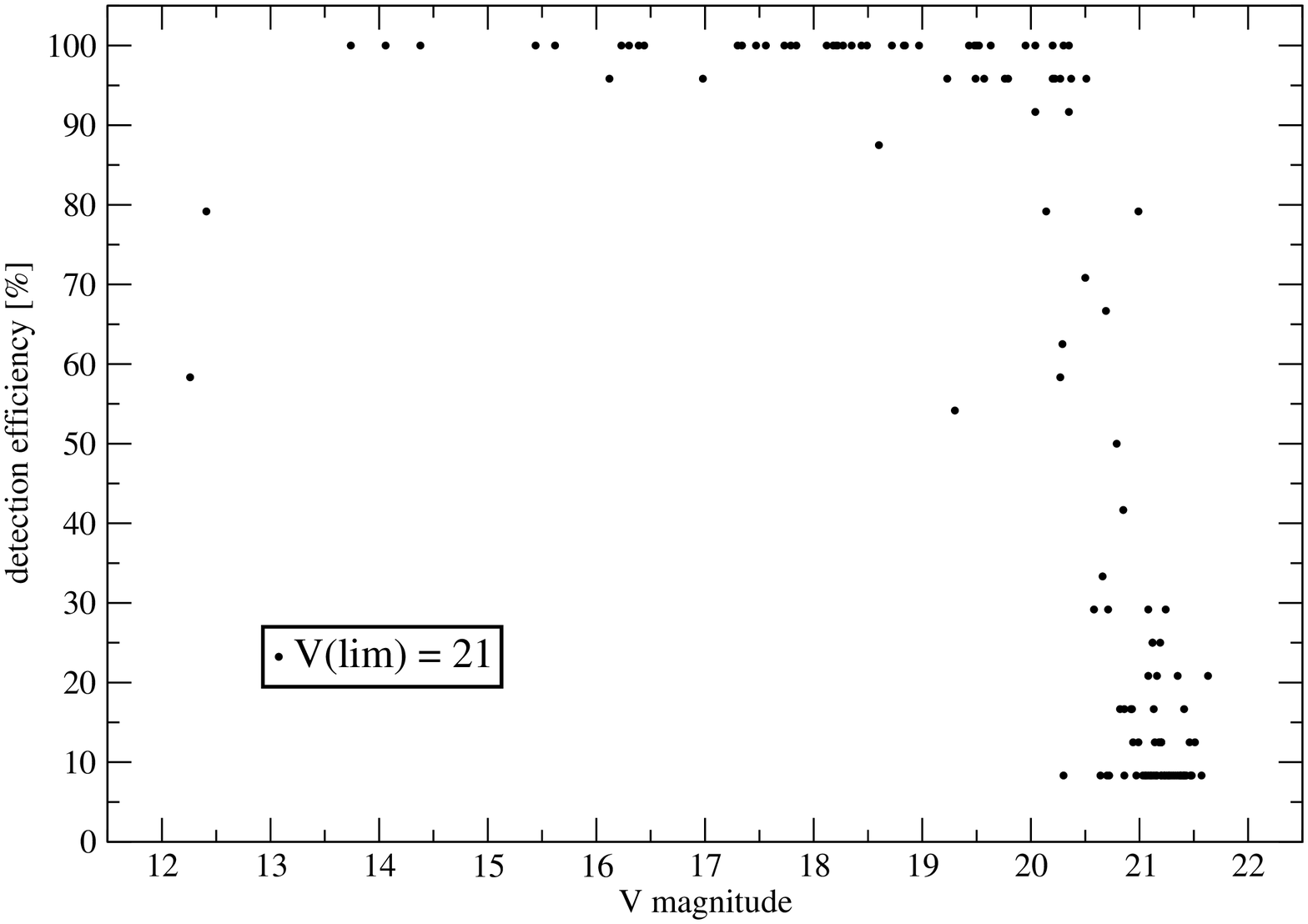}
\epsfysize=4.8cm 
\hspace{0.0cm} \vspace{0.0cm}
 \epsfbox{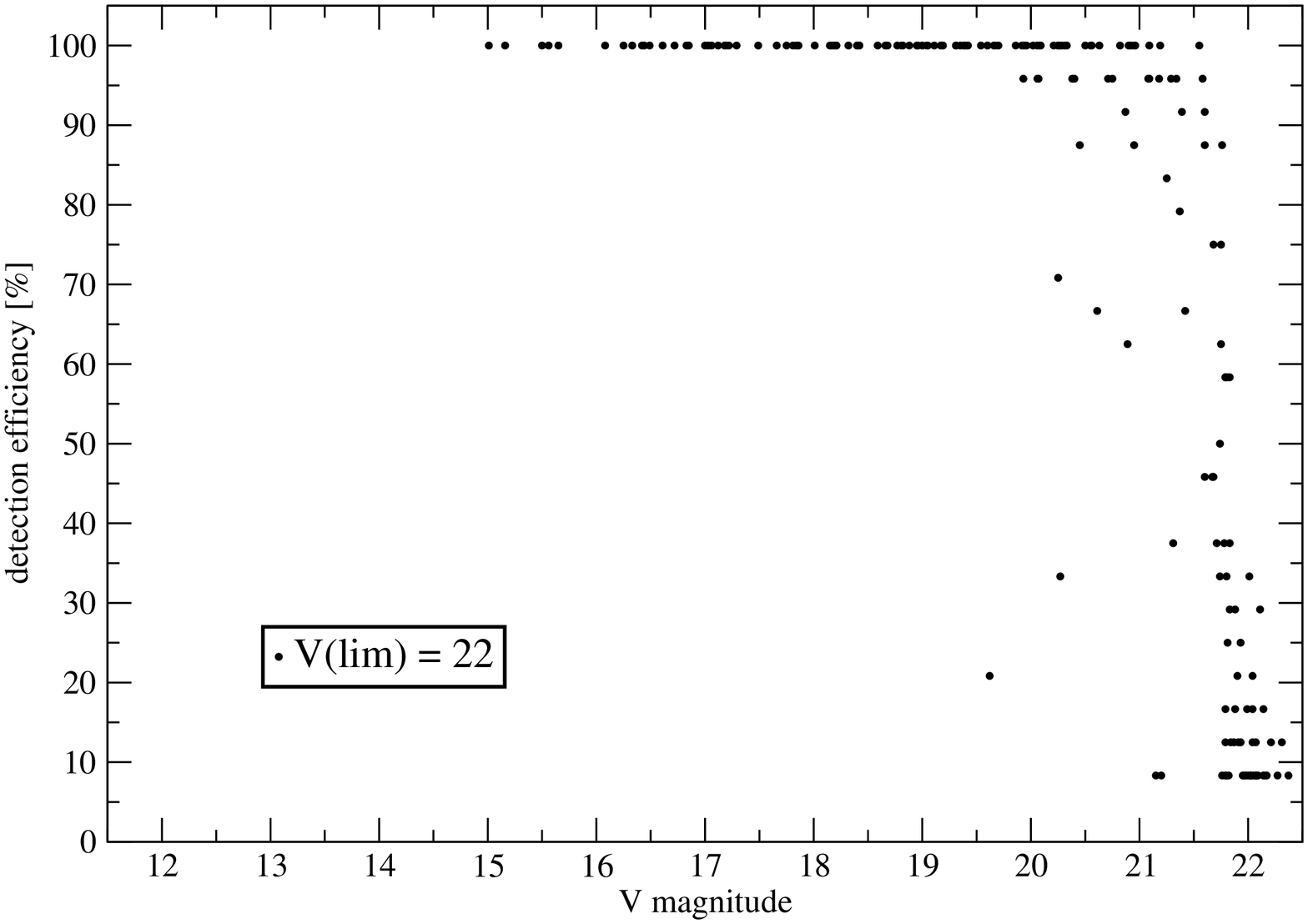}
\caption[h]{Two examples of detection efficiency as a function of $V$ magnitude.
Presented data are from pulsar fields \em PSR J 0056+4756 \em (left) and \em PSR
J 0454+5543 \em (right).
}
\end{figure}

We can conclude that a Crab like pulsar with an average $V$ magnitude of
22-24 would in the best possible scenario appear 1.8 magnitude brighter on every
$8^{th}$ image and so could be detected.

\subsection{Periodicity test}
As already mentioned, we observed in such a way that a pulsar was expected to
appear on every $8^{th}$ image in the set of 24. To look for such a periodic
signal we construct a set of 8 phase bins and compute the average magnitude
$\overline m^{\star}_i$ and its sigma $\sigma^{\star}_i$ for each bin. We divide
$\overline m^{\star}_i$ into two classes, those with higher and the ones with
lower magnitude than average. For each class we calculate average magnitude
values ($\overline m_{\uparrow}^{\star},\, \overline m_{\downarrow}^{\star}$)
and their sigmas ($\sigma_{\uparrow}^{\star},\, \sigma_{\downarrow}^{\star}$).
We define the periodicity measure as:
$$
S=\frac{\overline m_{\uparrow}^{\star}-\overline
m_{\downarrow}^{\star}}{\sqrt{{\sigma_{\uparrow}^{\star}}^2
+{\sigma_{\downarrow}^{\star}}^2}}\quad .
\eqno(2)
$$
Figure 5 presents the periodicity measure versus $V$ magnitude for
stars from two pulsar fields. Its values are scattered around the average value
of $\sim 0.7$, scattering is stronger for higher magnitudes. Each star with the
periodicity measure much over the average value is a possible candidate for the
pulsar optical counterpart only if its position agrees with the error box of the
radio pulsar position. On the sample of 12 pulsars no periodic signal that
would be a signature of a pulsar optical counterpart was detected.

\begin{figure}[h!]
\epsfysize=4.8cm 
\hspace{0.0cm} \vspace{0.0cm}
 \epsfbox{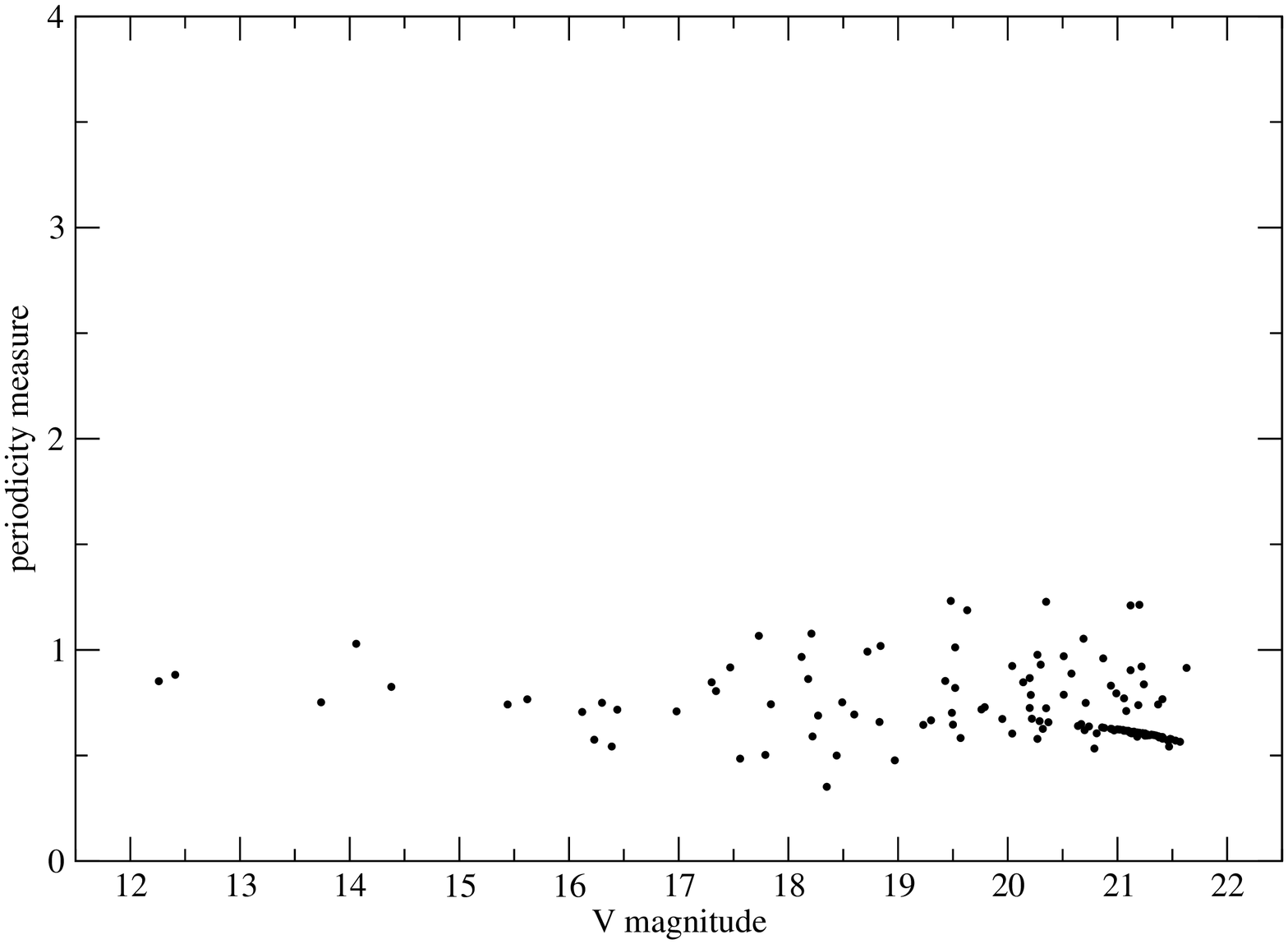}
\epsfysize=4.8cm 
\hspace{0.0cm} \vspace{0.0cm}
 \epsfbox{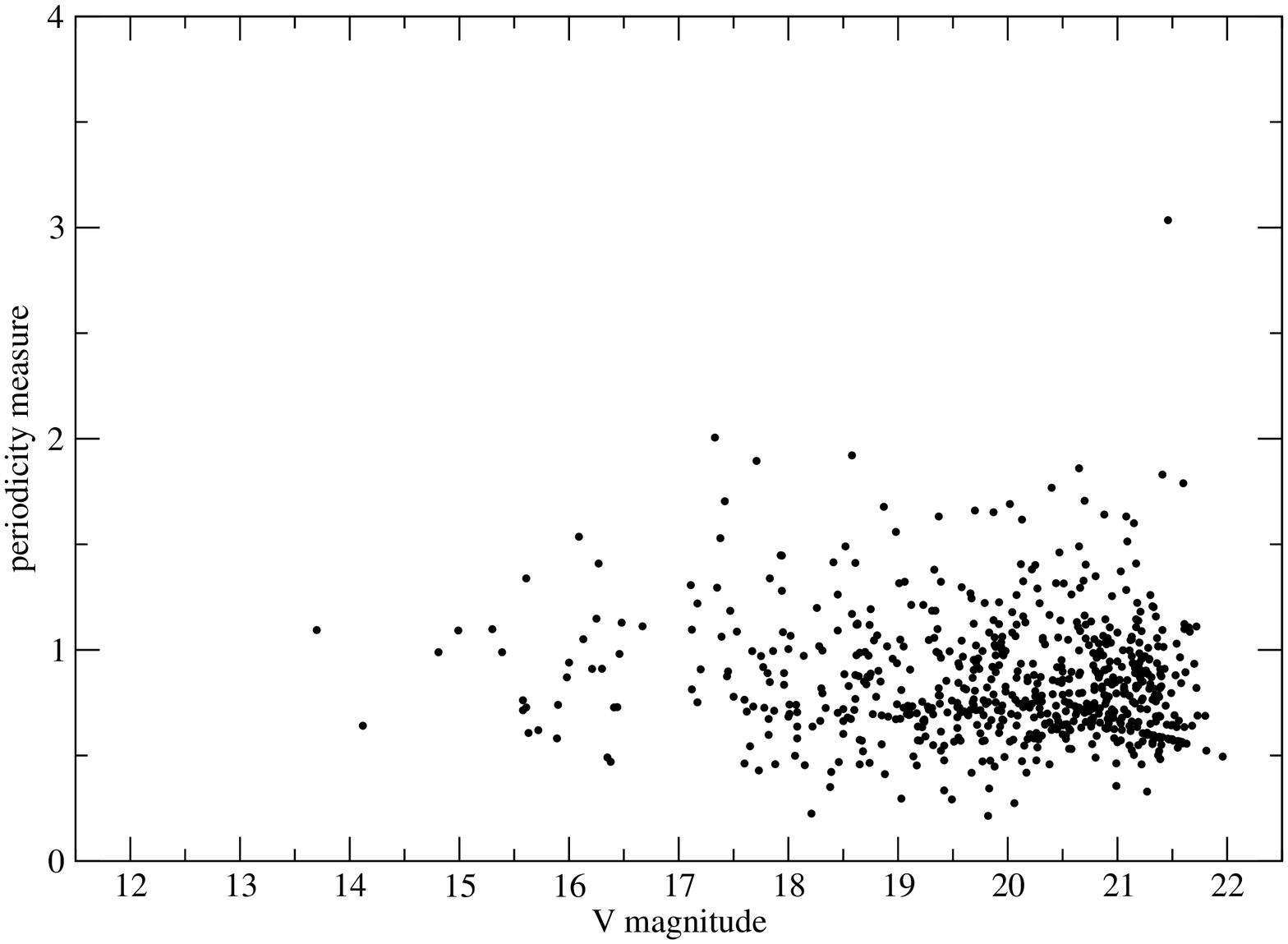}
\caption[h]{Two examples of periodicity measure versus $V$
magnitude. Presented data are from pulsar fields \em PSR J 0056+4756 \em (left)
and \em PSR J 1833-0827 \em (right).
}
\end{figure}

\section{Conclusion}
With the analysis described above no optical counterpart with an average
magnitude of 22-24 on the sample of 12 radio pulsars was detected.
Optical counterparts already discovered are very faint. The limiting magnitude
that we reached with the Clypos survey is most probably not high enough for any
new detection. However, the stroboscopic method is a useful observational
technique since it enables not only detection but also periodicity
identification of optical counterparts. Once the absolute phase of the pulsar is
known one can synchronize the stroboscope and thus fully use the reduction of
the sky brightness contribution resulting from magnitude enhancement. For
example, with the VLT using active optics and in stroboscopic mode we would
expect that 0.01 photometry of a 25 average magnitude pulsar would be allowed in
10 minutes.

\acknowledgements
We thank Dick Manchester and David Nice for help with the TEMPO program and the
technical staff of the Guillermo Haro Observatory for professional support and
friendly atmosphere. This program was supported in part by the Ministry of
Education, Science and Sport of the Republic of Slovenia and in part by CONACyT
grant 25539E.


\end{document}